\documentclass[namedreferences,hyperref,optionalrh]{spr-sola}
\usepackage{graphicx}
\usepackage{amssymb}
\usepackage{color}
            
\usepackage{bm}
\usepackage{amsmath}

\usepackage[]{hyperref}
\hypersetup{
    colorlinks=true,
    urlcolor=blue,
    citecolor=blue,
    linkcolor=blue,
}

\chardef\us=`\_

\begin{document}

\begin{frontmatter}

\title{The Correlation Length of Turbulence in Magnetic Clouds}

\author[addressref=aff1,corref,email={simon.good@helsinki.fi}]{\inits{S.W.}\fnm{S.W.}~\snm{Good}\orcid{0000-0002-4921-4208}}
\author[addressref=aff2,email={jlalu002@gmail.com}]{\inits{J.}\fnm{J.}~\snm{Lalueza Pu\'ertolas}\orcid{0009-0004-8988-6596}}
\author[addressref={aff3},email={jylhae@ualberta.ca}]{\inits{A.-S.M.}\fnm{A.-S.M.}~\snm{Jylhä}\orcid{0000-0002-2918-5120}}
\author[addressref=aff1,email={emilia.kilpua@helsinki.fi}]{\inits{E.K.J.}\fnm{E.K.J.}~\snm{Kilpua}\orcid{0000-0002-4489-8073}}

\address[id=aff1]{Department of Physics, University of Helsinki, PO Box 64, 00014 Helsinki, Finland}
\address[id=aff2]{Faculty of Physics, University of Barcelona, Mart\'i i Franqu\`es 1-11, 08028 Barcelona, Spain}
\address[id=aff3]{Department of Physics, University of Alberta, Edmonton AB T6G 2E1, Canada}

\runningauthor{Good et al.}
\runningtitle{Turbulence Correlation Length in Magnetic Clouds}

\begin{abstract}
The large-scale limit or outer scale of turbulence in the solar wind is associated with the correlation length of the magnetic field. Determining correlation lengths from magnetic field time series in magnetic clouds is complicated by the presence of the global flux rope: without removal of the flux rope trend, correlation length measurements will be sensitive to the flux rope as well as the turbulence, and give overestimates of the outer scale when turbulence amplitudes at the outer scale are small relative to the flux rope amplitude. We have used force-free flux rope fits to detrend magnetic field time series measured by \textit{Parker Solar Probe} in two magnetic clouds and calculated the turbulence correlation length in the clouds using the detrended data. The detrended correlation length in terms of the proton inertial length, $d_p$, was $2.7\times10^{4} d_p$ in one cloud (observed at 0.77~au) and $1.6\times10^{4}d_p$ in the other (observed at 0.39~au), significantly smaller than the values obtained without detrending. Increments in the flux rope fits scaled equivalently to a $k^{-3}$ wavenumber power spectrum; this contribution from the flux rope considerably steepened the total spectrum at the largest scales but had a negligible effect in the inertial range, where scaling in both clouds equivalent to $\sim$$k^{-5/3}$ was observed. Finally, we discuss the possible relation of turbulence correlation lengths to mesoscale structure in magnetic clouds. 

\end{abstract}
\end{frontmatter}

\section{Introduction}
     \label{sec:intro} 

The solar wind is a turbulent mix of plasma streams and structures flowing from the Sun to the heliopause. The most significant perturbations to the quasi-steady outflow of the solar wind are caused by the eruption of large-scale coronal flux ropes, which, along with their associated effects, are observed as interplanetary coronal mass ejections \citep[ICMEs;][]{Kilpua17} by spacecraft in situ. ICMEs are termed magnetic clouds when they display a specific set of ideal signatures in spacecraft time series measurements, namely the global, monotonic rotation of the magnetic field vector indicating the passage of a flux rope, low proton temperatures and an associated low proton plasma beta ($\beta_p\ll1$), and enhanced magnetic field magnitudes relative to the ambient solar wind \citep{Burlaga81,Zurbuchen06}. ICME flux ropes can carry sustained periods of southward magnetic field to the Earth, making them a major driver of space weather \citep[e.g.][]{Zhang07}.

A key objective of heliophysics research is to achieve a better understanding of ICME structure in order to improve the accuracy of space weather forecasting models and to better inform choices being made about future space weather missions \citep[e.g.][]{Lugaz25,Palmerio25b,Weiler25}. While the basic paradigm of the ICME driver as a large-scale, twisted flux tube (i.e. flux rope) originating from the Sun is well supported by theory and observations \citep[e.g.][]{Bothmer98}, the true variability of ICME properties is not accommodated by simplified global models \citep{AlHaddad25}. An important feature absent from the simple models has been highlighted by \citet{Owens20}, who demonstrated how different regions within an ICME can expand away from each other faster than Alfv\'en waves (i.e. information) can propagate between them, and that ICMEs thus cease to behave as globally coherent magnetohydrodynamic (MHD) structures at certain sizes and distances from the Sun; ICMEs will distort locally to interactions rather than respond like rigid bodies. Localised distortion could partly explain ICME `complexity', which describes non-ideal features of ICME structure that can vary significantly between spacecraft observing the same ICME even at relatively small angular or radial separations \citep[e.g.][]{Kilpua11,Scolini22,Banu25}. Besides being distortions of an initially simple structure that gradually develop with propagation from the Sun, complex features could also be more intrinsic to ICME flux ropes and exist at the earliest stages of their development in the corona \citep[e.g.][]{Daei23}. 

One phenomenon that has not been examined in detail as a potential source of ICME complexity is MHD turbulence. Like the solar wind in general, ICME plasma displays a broadband spectrum of fluctuations consistent with the presence of a turbulent cascade \citep[e.g. as recently studied by][]{Good20a,Good23,Bhattacharjee23,Shaikh24,Rakhmanova25,Bai25,Ruohotie25,Khuntia25}. A characteristic length scale of turbulent systems is the `outer scale' at which energy is injected and from which the energy cascades to successively smaller scales. The outer scale is associated with the correlation length of the fluctuations in a turbulent system \citep[e.g.][]{Matthaeus82}. In the case of solar wind parcels advecting supersonically past an observing spacecraft, the correlation time can be estimated from the mean-subtracted or detrended magnetic field time series measured through the parcel and converted to a length scale with Taylor's hypothesis. Correlation lengths vary with heliocentric distance \citep{Ruiz14}, solar wind type \citep{Isaacs15} and the solar cycle \citep{Wicks10}.

Correlations in the magnetic fields of ICMEs have been examined in a number of studies \citep[e.g.][]{Good18,Lugaz18}. \citet{Lugaz18} determined the Pearson correlations between field profiles observed by the \textit{Wind} and \textit{Advanced Composition Explorer} spacecraft at a range of non-radial separation distances within 35 ICMEs, finding the zero-correlation separation distance via extrapolation to be $\sim$0.1~au for the field components and $\sim$0.3~au for the field magnitude. Here we note an important distinction between these length scales and the correlation length of turbulence as described previously. Lugaz et al. calculated correlations in 30-min averaged data in order to sample the global magnetic structure (i.e. background trends) within the ICME intervals; conversely, turbulence correlation lengths are calculated from the residual fluctuations, which are obtained by removing such background trends from the data. \citet{Ruiz14} estimated correlation lengths in ICMEs at 1~au with linearly detrended data, finding the mean length in ICMEs to be significantly longer than in the solar wind generally. Recently, \citet{Good23} calculated turbulence correlation lengths in 15 magnetic clouds observed in the inner heliosphere by detrending the time series with non-linear flux rope fits.

In this work, correlation lengths in magnetic clouds are explored in more detail. Two clouds with relatively simple large-scale flux rope structures but with significantly different turbulence amplitudes have been selected for analysis. In Section~\ref{sec:rope_fits}, fits to the magnetic field profiles of the clouds using standard flux rope models are described. These fits have been used to detrend the profiles in order to calculate correlation lengths of the turbulence in the clouds, as described in Section~\ref{sec:corr_length}. Without detrending, we show how correlation lengths will measure large-scale correlations of the background flux rope when turbulence amplitudes at the outer scale are small relative to the flux rope amplitude, as is usually the case in magnetic clouds. If turbulence amplitudes are large, by contrast, correlation lengths calculated without detrending will be less sensitive to the flux rope, more sensitive to the turbulence and hence shorter. In Section~\ref{sec:B_rot}, distributions of fluctuation amplitudes across a range of fluctuation timescales in the two clouds are examined. In the total fields, the mean values of these distributions show power-law scalings expected of turbulent fluctuations below the detrended correlation lengths and steeper scalings above them, where the flux rope field dominates. The significance of correlation length measurements for understanding turbulence and `mesoscale structure' in ICMEs is briefly discussed in Section~\ref{sec:discussion}.

\section{Analysis}
\label{sec:analysis}

\begin{figure}
\centerline{\includegraphics[width=1.05\textwidth,clip=]{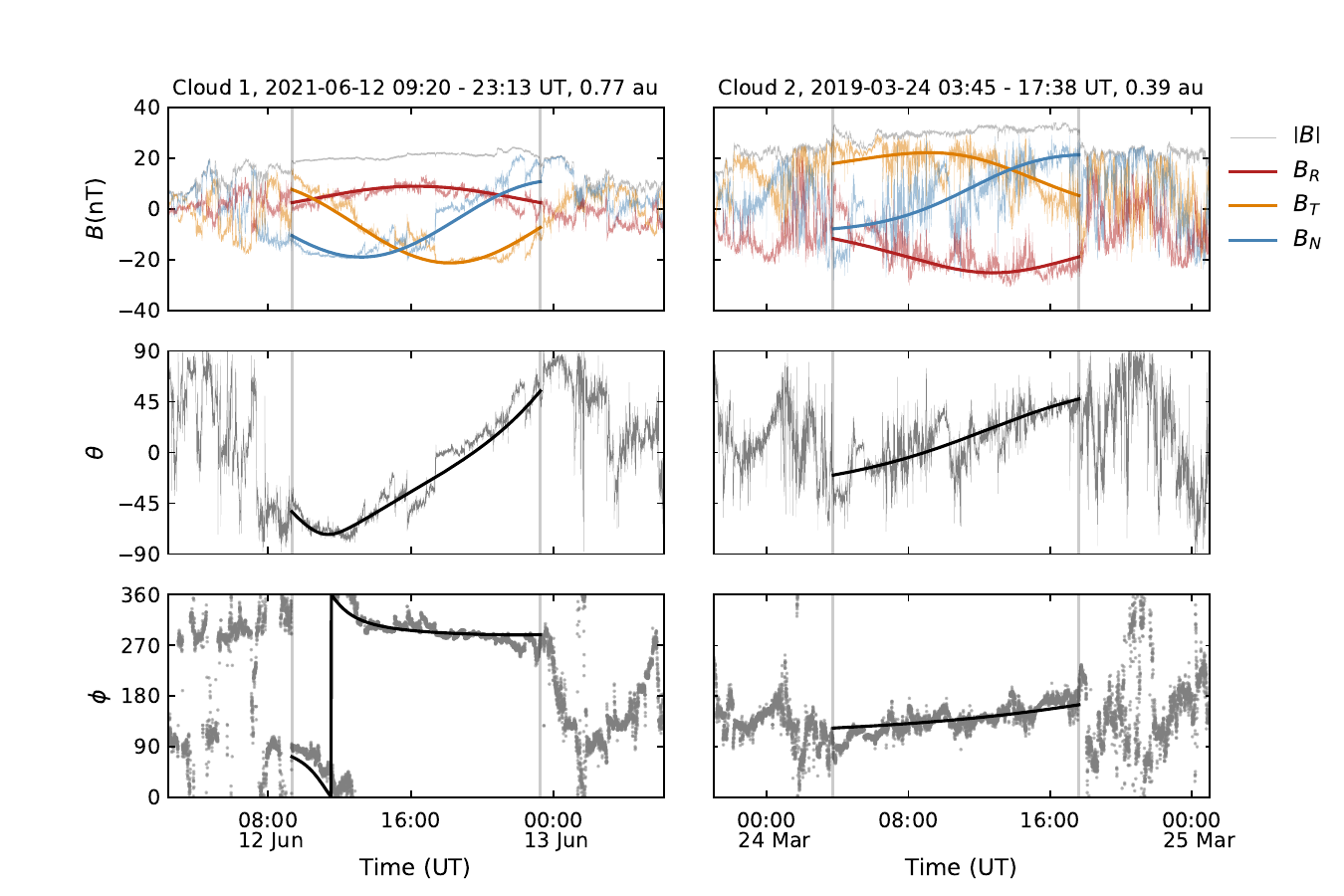}}
\small
        \caption{Magnetic cloud observations made by PSP in June 2021 at 0.77~au (left-hand panels) and March 2019 at 0.39~au (right-hand panels). The cloud intervals are bounded by vertical lines. From top to bottom, the panels show the magnetic field magnitude and components in RTN coordinates, the latitude angle, $\theta$, of the magnetic field vector relative to the $R$-$T$ plane, and longitude angle, $\phi$, between the $R$ direction and the projection of the magnetic field vector onto the $R$-$T$ plane. Smooth lines show flux rope fits to the data.}
\label{fig:events}
\end{figure}

Figure~\ref{fig:events} displays magnetic field measurements made by the FIELDS instrument suite \citep{Bale16} on board \textit{Parker Solar Probe} \citep[PSP;][]{Fox16} during the passage of two magnetic clouds over the spacecraft. Data are shown in radial-tangential-normal (RTN) coordinates at 0.11~s resolution. The elevated magnetic field magnitude and global, monotonic field rotation characteristic of magnetic clouds are evident in both cloud intervals, which were both of duration $T=13.9$~hours. A significant difference between the two clouds can be seen in the amplitudes of small-scale magnetic field fluctuations, which are visibly smaller in the June 2021 cloud (`cloud 1'; left-hand panels) than in the March 2019 cloud (`cloud 2'; right-hand panels). Cloud 1 had a mean proton plasma beta $\langle \beta_p \rangle=0.05$, mean proton bulk speed $\langle v_p \rangle=$~404~km~s$^{-1}$ and radial width $L=\langle v_p \rangle T=2.02\times10^{7}$~km = 0.135~au, while cloud 2 had $\langle \beta_p \rangle=0.08$, $\langle v_p \rangle=$~327~km~s$^{-1}$ and $L=1.64\times10^{7}$~km = 0.109~au. These and other key parameters relating to the two clouds are listed in Table~\ref{tab:parameters}.

\begin{table}
\caption{Measured and derived parameters relating to two magnetic clouds observed by PSP.}
\label{tab:parameters}
\begin{tabular}{lll}

\hline
Parameter & Cloud 1 & Cloud 2 \\
\hline
\textit{Key parameters} &  &  \\
Heliocentric distance, $r$ (au) & 0.77 & 0.39 \\
Duration, $T$ (hr) & 13.9 & 13.9   \\
Proton plasma beta, $\langle\beta_p\rangle$ & 0.05 & 0.08  \\
Proton bulk speed, $\langle v_p\rangle$ (km~s$^{-1}$) & 404 & 327 \\
Proton inertial length, $\langle d_p \rangle$ (km) & 54 & 29 \\
Radial width, $L=\langle v_p\rangle T$ (km) & $2.02\times10^{7}$ & $1.64\times10^{7}$ \\
\hline
\textit{Flux rope fitting} &  &  \\
Axis magnetic field magnitude, $B_0$ (nT) & 25.4 & 36.7 \\
Handedness, $H$ & $-1$ & -- \\
Field-line twist, $\Gamma$ (au$^{-1}$) & -- & 3.0 \\
RTN axis latitude angle, $\theta_0$ ($^\circ$) & $-35$ & 31 \\
RTN axis longitude angle, $\phi_0$ ($^\circ$) & 280 & 119 \\
Impact parameter, $p$ & 0.19 & 0.58 \\
\hline
\textit{Correlation scales} &  &  \\
Correlation time (mean-subtracted), $\tau_c$ (min) & 163 & 42 \\
Correlation time (rope-subtracted), $\tau'_c$ (min) & 59 & 24 \\
Turbulence correlation length, $L'_c=\langle v_p\rangle\tau'_c $ (km) & $1.43\times10^{6}$ & $0.47\times10^{6}$ \\
$L'_c/L$ & 0.071 & 0.029 \\
$L'_c/\langle d_p \rangle$ & $2.65\times10^{4}$ & $1.61\times10^{4}$ \\
\hline
\end{tabular}
\end{table}

\subsection{Flux Rope Fitting}
      \label{sec:rope_fits}  

Flux rope fits to the magnetic field time series are shown by smooth lines in Figure~\ref{fig:events}. Cloud 1 has been fitted with the force-free \citet{Lundquist51} solution, described in cylindrical coordinates by 
\begin{equation}
\bm{B}(\rho) = B_0 J_0(\alpha\rho) \bm{\hat{z}} + H B_0 J_1(\alpha\rho) \bm{\hat{\phi}},
\label{eq:Lundquist}
\end{equation}
where $\rho$, $z$ and $\phi$ are the radial, axial and tangential coordinates, respectively, $J_0$ and $J_1$ are the zeroth and first-order Bessel functions, respectively, $B_0$ is the magnetic field magnitude at the axis, $H=\pm1$ is the rope handedness and $\alpha$ is the force-free parameter, with $\alpha$ here assumed to be constant. The rope radius, $\rho_0$, is set by convention to the first zero of $J_0$ such that $\alpha\rho_0\simeq2.405$. Besides $B_0$ and $H$, key parameters obtained from the fitting include the latitude angle, $\theta_0$, and longitude angle, $\phi_0$, of the flux rope axis relative to the $R$-$T$ plane, and the distance of closest approach made by the spacecraft to the rope axis normalised to the rope radius, $p$. Lundquist fit parameters obtained for cloud 1 are: $B_0=25.4$~nT; $H=-1$; $\theta_0=-35^\circ$; $\phi_0=280^\circ$; and $p=0.19$. Further details of the fitting procedure are described by \citet{Good19}.

The large-angle rotation executed by the magnetic field in cloud 1 is well captured by the Lundquist fit. The smaller global angle of rotation seen in cloud 2, by contrast, is more accurately reproduced with the force-free \citet{GoldHoyle60} solution, given by 
\begin{equation}
\bm{B}(\rho) = \frac{B_0}{1 + \Gamma^2 \rho^2}\bm{\hat{z}} + \frac{B_0\Gamma\rho}{1 + \Gamma^2 \rho^2}\bm{\hat{\phi}},
\label{eq:GoldHoyle}
\end{equation}
where $\Gamma$ is the constant field-line twist, defined as the number of complete turns made by field lines per unit length in the $z$ direction. The rope radius is fixed at $\rho_0=1/\Gamma$. Unlike in the uniformly twisted Gold-Hoyle rope, twist in the Lundquist rope varies with $\rho$, being at a minimum at the axis and infinite at the boundaries. Gold-Hoyle fit parameters obtained for cloud 2 are: $B_0=36.7$~nT; $\Gamma=3.0$~au$^{-1}$; $\theta_0=31^\circ$; $\phi_0=119^\circ$; and $p=0.58$. These fitting results were previously reported by \citet{Good23}.

\subsection{Turbulence Correlation Length}
      \label{sec:corr_length}      

\begin{figure}
\centerline{\includegraphics[width=\textwidth,clip=]{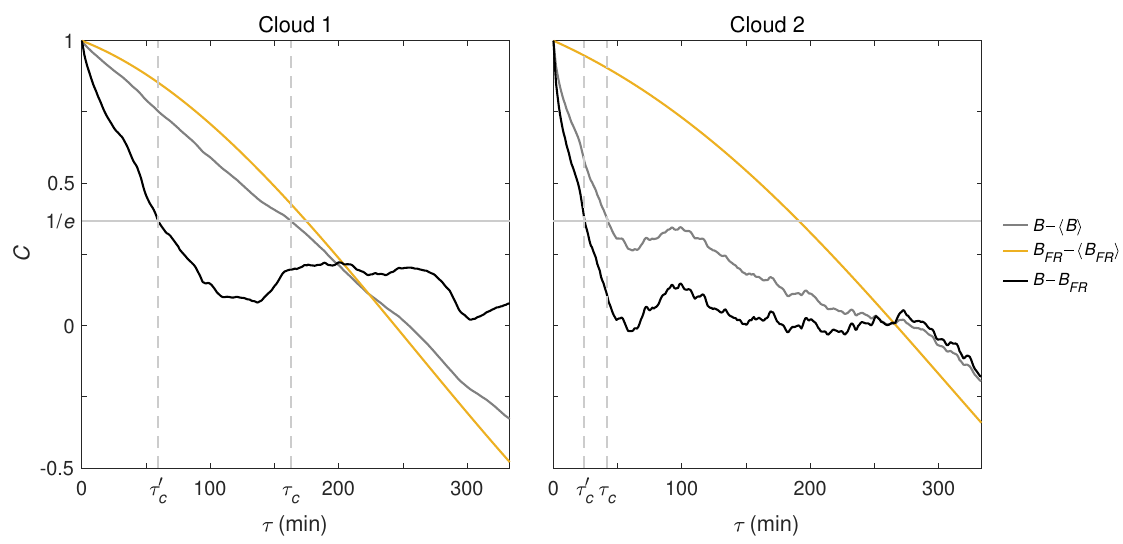}}
\small
        \caption{Autocorrelation curves as functions of time lag for the mean-subtracted magnetic field (grey curves), fit-subtracted magnetic field  (black curves) and mean-subtracted flux-rope fit (gold curves) in each magnetic cloud. The grey curves are sensitive to the rotation of the flux rope field and to small-scale turbulent fluctuations, while the black curves are sensitive to the turbulence only. The $1/e$ correlation times in the mean-subtracted fields, $\tau_c$, and fit-subtracted fields, $\tau'_c$, are marked by vertical dashed lines.}
\label{fig:correlations}
\end{figure}

The correlation time of turbulent fluctuations in a magnetic field time series, $\bm{B}(t)$, may be obtained from the normalised autocorrelation function,
\begin{equation}
    C(\tau) = \frac{\left\langle \delta\bm{B}(t) \cdot \delta\bm{B}(t+\tau) \right\rangle}{\left\langle|\delta\bm{B}(t)|^2\right\rangle},
    \label{eqn:autocorr}
\end{equation}
where ${\delta\bm{B}(t) = \bm{B}(t) - \bm{B}_G}$ is calculated by subtracting an estimate of the background field, $\bm{B}_G$, from the measured time series, where $\tau$ is a time lag, and where angle brackets denote time averaging over a selected interval. The correlation time, $\tau_c$, is here defined such that ${C(\tau_c)=1/e}$. Autocorrelation curves have been calculated with intervals of duration \mbox{$\chi=500$~min} (equal to $0.6T$ in both clouds) and $\tau$ ranging from 0 to 333~min ($0.4T$). The condition ${\chi+\tau\leq T}$ was imposed so that the trailing solar wind was not sampled in the $C(\tau)$ calculation.

\overfullrule=0pt
The grey curves in Figure~\ref{fig:correlations} show $C(\tau)$ calculated with constant ${\bm{B}_G=\left\langle \bm{B}(t) \right\rangle_T}$ values, where the time average is taken across the entire cloud interval in each instance. The black curves show $C(\tau)$ calculated with ${\bm{B}_G=\bm{B}_{FR}}$, where $\bm{B}_{FR}(t)$ are the time series of the flux rope fits described in Section~\ref{sec:rope_fits}; $\delta\bm{B}(t)$ is thus obtained by a point-by-point subtraction of the fit values from the measured time series. The gold curves show $C(\tau)$ for the fit time series and calculated in the same way as for the grey curves, i.e. with ${\delta\bm{B}(t) = \bm{B}_{FR}(t) - \left\langle \bm{B}_{FR}(t) \right\rangle_T}$. 

For cloud 1, the grey curve follows closely the autocorrelation of the fit (gold curve): the large-amplitude, low-frequency rotation of the flux rope has not been removed by a simple subtraction of the mean field and remains as a dominant component of $\delta\bm{B}(t)$ that determines the overall correlation. In this case, the small-scale turbulent fluctuations only cause a minor deviation of the overall correlation from that of the flux rope field, represented by the gold curve. The flux rope field is correlated over a relatively long time period, hence the relatively long correlation time, \mbox{$\tau_c=163$~min}, obtained from the rope-dependent grey curve. The increasingly negative values of the grey and gold curves at large $\tau$ reflect systematic anti-correlations between different segments of the global flux rope field. In contrast, the black curve gives a much shorter correlation time of \mbox{$\tau'_c=59$~min}; the influence of the flux rope rotation has been removed from the calculation in this case, and $\tau'_c$ represents the correlation time of the residual turbulent fluctuations alone. Moreover, the exponentially decaying form of the black curve is consistent with the autocorrelation behaviour expected of turbulence \citep{Matthaeus82}.

The corresponding trends are less pronounced for cloud 2. Here the turbulent fluctuations are closer in amplitude to that of the flux rope and thus have a much larger relative impact on the grey curve, which deviates significantly from the fit autocorrelation. The black curve obtained from the rope-subtracted field is similar in form to the rope-dependent grey curve, although the correlation time is still reduced with rope subtraction; \mbox{$\tau_c=42$~min} for the grey curve and \mbox{$\tau'_c=24$~min} for the black curve in cloud 2.

Correlation lengths provide a more physically significant parameter than correlation times when making comparisons between different intervals. Applying Taylor's hypothesis, the turbulent (i.e. rope-subtracted) correlation length, ${L'_c= \langle v_p \rangle\tau'_c}$, for clouds 1 and 2 are \mbox{$1.43\times10^{6}$~km} and \mbox{$0.47\times10^{6}$~km}, respectively, i.e. \mbox{$\sim$3} times larger in cloud 1 than in cloud 2. This factor reduces to \mbox{$\sim$2.4} when comparing the correlation lengths normalised to the clouds' radial widths, ${L'_c/L}$, which are 0.071 and 0.029 in clouds 1 and 2, respectively, and further reduces to \mbox{$\sim$1.6} when the correlation lengths are expressed in terms of the mean proton inertial lengths, $\langle d_p \rangle$; ${L'_c=2.65\times10^4\langle d_p \rangle}$ in cloud 1 and ${L'_c=1.61\times10^4\langle d_p \rangle}$ in cloud 2. The correlation lengths obtained without the flux rope fields subtracted, ${L_c= \langle v_p \rangle\tau_c}$, are $3.95\times10^{6}~\text{km}=7.33\times10^4\langle d_p \rangle$ in cloud 1 and $0.82\times10^{6}~\text{km}=2.83\times10^4\langle d_p \rangle$ in cloud 2, significantly longer than the corresponding ${L'_c}$ values.

\subsection{Magnetic Field Fluctuations}
      \label{sec:B_rot}    

\begin{figure}
\centerline{\includegraphics[width=1.02\textwidth,clip=]{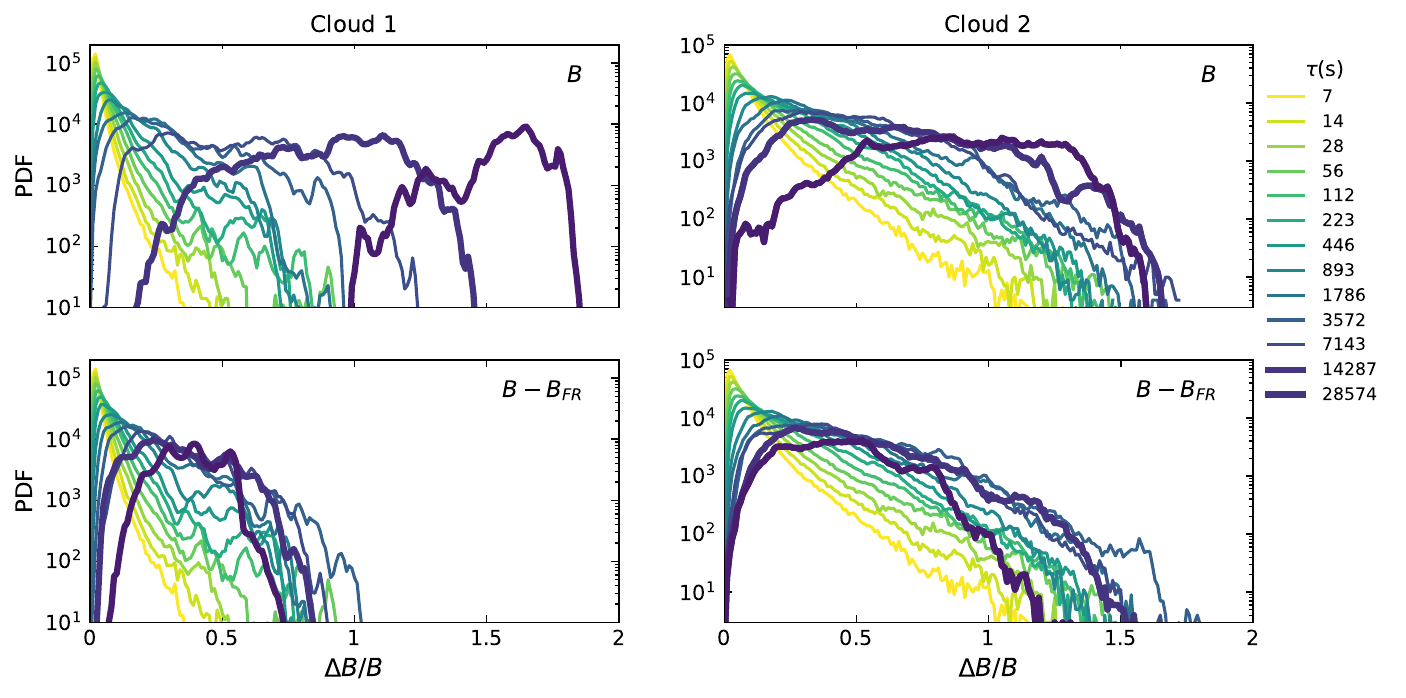}}
\small
        \caption{PDFs of magnetic field fluctuation amplitudes normalised to the local mean field magnitude as functions of fluctuation timescale. Top and bottom panels show distributions obtained with the total and rope-subtracted field, respectively.}
\label{fig:distributions}
\end{figure}      

\sloppy
Amplitudes of magnetic field fluctuations across a range of timescales, $\tau$, have also been examined in the two clouds. The top panels in Figure~\ref{fig:distributions} show distributions of fluctuation amplitudes ${\Delta B_{tot}(t,\tau)=|\bm{B}(t+\tau)-\bm{B}(t)|}$ for the total magnetic field, $\bm{B}(t)$, where each amplitude has been normalised to the mean of the field magnitudes at times $t$ and $t+\tau$ (i.e. the `local magnitude') such that
\begin{equation}
    \Delta B_{tot}/B = 2\frac{|\bm{B}(t+\tau)-\bm{B}(t)|}{|\bm{B}(t+\tau)|+|\bm{B}(t)|}.
    \label{eqn:norm_fluc1}
\end{equation}
Distributions are displayed for successively doubled values of $\tau$ in the range \mbox{7--28~574~s}. The bottom panels show the corresponding distributions for the rope-subtracted field, ${\delta\bm{B}(t)=\bm{B}(t)-\bm{B}_{FR}(t)}$, where fluctuation amplitudes ${\Delta B'(t,\tau)=|\delta\bm{B}(t+\tau)-\delta\bm{B}(t)|}$ are normalised in the same way as Equation~\ref{eqn:norm_fluc1} to give
\begin{equation}
    \Delta B'\!/B = 2\frac{|\delta\bm{B}(t+\tau)-\delta\bm{B}(t)|}{|\bm{B}(t+\tau)|+|\bm{B}(t)|}.
    \label{eqn:norm_fluc2}
\end{equation}
In the incompressible limit, the field magnitude is constant (i.e. $|\bm{B}(t+\tau)|=|\bm{B}(t)|$), fluctuations $\Delta B$ (i.e. $\Delta B_{tot}$ or $\Delta B'$) are the amplitudes of pure rotations made by the field vector on a sphere with radius $B$, and the maximum possible rotation of $180^\circ$ imposes the limit $\Delta B/B\leq 2$. Furthermore, by defining an equivalent amplitude of increments in the flux rope field, ${\Delta B_{FR}(t,\tau)=|\bm{B}_{FR}(t+\tau)-\bm{B}_{FR}(t)|}$, it may be shown that
\begin{equation}
    \Delta B_{tot}(t,\tau)=\Delta B_{FR}(t,\tau)+\Delta B'(t,\tau),
    \label{eqn:amp_sum}
\end{equation}
which expresses the total fluctuation at every time and scale as the sum of a flux rope and non-flux rope (i.e. turbulent) component.

\begin{figure}
\centerline{\includegraphics[width=1.05\textwidth,clip=]{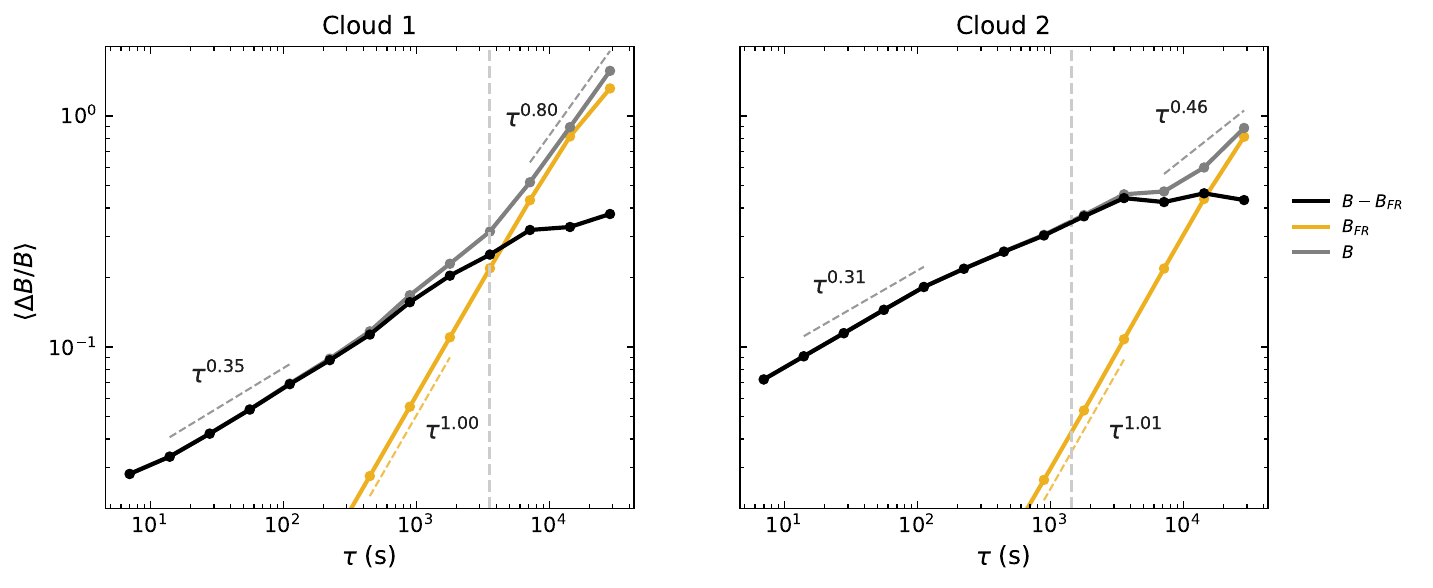}}
\small
        \caption{Mean values of the PDFs in Figure~\ref{fig:distributions}  as functions of fluctuation timescale for the total field (grey lines) and rope-subtracted field (black lines). Also shown are the mean values of the equivalent PDFs calculated from the fitted flux rope fields (gold lines). Vertical dashed grey lines mark the rope-subtracted correlation timescales, $\tau'_c$, obtained from the autocorrelation curves shown in Figure~\ref{fig:correlations}.}
\label{fig:spectra}
\end{figure}

In Figure~\ref{fig:distributions}, distributions at small $\tau$ peak sharply at small fluctuation amplitudes and have long tails extending to relatively high amplitudes, indicative of the field vector typically rotating by small angles (and thus being well autocorrelated) over short timescales, but having a small though statistically significant probability of undergoing large rotations. As $\tau$ increases, the mean amplitude increases, the tails diminish and the distributions become more Gaussian; i.e., rotation angles increase and the vector autocorrelation decreases. These qualitative trends are present at $\tau\lesssim10^3\,\text{s}$ in both clouds with and without rope subtraction (though scaled up to larger amplitudes in cloud 2), and can be explained by MHD turbulence models \citep{Zhdankin12}. The distribution tails may be interpreted as turbulent intermittency, a phenomenon whereby energy cascading to smaller scales becomes increasingly concentrated in fluctuations with large amplitudes relative to the mean amplitude at each scale; it has also been suggested that such tails represent flux tube boundaries \citep{Borovsky08}. Similar trends in $\Delta B/B$ distributions and the equivalent rotation angle distributions have been seen generally in the solar wind \citep{Zhdankin12,Chen15,Matteini18,Larosa24} and in ICME sheaths \citep{Good20b,Kilpua20}.

In their study of the fast solar wind at 1.4--2.2~au, \citet{Matteini18} found that $\Delta B/B$ distributions became roughly symmetric at $\tau\simeq5\times10^3\,\text{s}$ and did not evolve with further increases of scale, settling at a constant mean value $\langle\Delta B/B\rangle\simeq1$. This behaviour reflects the presence of the well-known $1/f$ spectrum of non-turbulent, large-amplitude fluctuations in the fast wind at scales exceeding the outer scale. The distributions for the total fields in the two clouds (Figure~\ref{fig:distributions}, top panels) display a different behaviour, particularly in cloud 1. Rather than settling at a constant value at $\tau\gtrsim10^3\,\text{s}$, $\langle\Delta B_{tot}/B\rangle$ continues to grow with $\tau$ due to the flux rope rotation, which is the dominant component of the field at these large timescales; flux rope fields trivially rotate by larger angles over larger time increments, with $\Delta B_{tot}(0,T)/B\simeq2$ for a $\sim180^{\circ}$ rotation between the leading and trailing edge of a flux rope. In cloud 2, $\langle\Delta B_{tot}/B\rangle$ does not reach the high values at large $\tau$ seen in cloud 1 because of the smaller global rotation of the rope in cloud 2. With subtraction of the flux rope field (Figure~\ref{fig:distributions}, bottom panels), the distributions at large $\tau$ are qualitatively similar to the fast wind case, but with $\langle\Delta B'\!/B\rangle<1$.

The trends described above are also evident in Figure~\ref{fig:spectra}, which shows $\langle\Delta B_{tot}/B\rangle$ (grey lines) and $\langle\Delta B'\!/B\rangle$ (black lines) plotted versus $\tau$. The trends take power-law forms, i.e. $\langle\Delta B/B\rangle\propto\tau^{\alpha_\tau}$. As can be seen in Figure~\ref{fig:events}, the local field magnitude is approximately scale-independent in the two clouds and $\langle \Delta B/B\rangle$ is equivalent to the scale-dependent $\langle \Delta B\rangle$ normalised to a constant $B$ \citep{Matteini18}. Therefore, $\langle\Delta B/B\rangle\propto\langle\Delta B\rangle$ and the scalings in Figure~\ref{fig:spectra} can be directly and straightforwardly related to the wavenumber power spectrum, $P(k)\propto k^{\alpha_k}$, via the relation $\Delta B^2=P(k)k$ \citep[e.g.][]{Monin75} and Taylor's hypothesis, the latter giving $k\propto1/\tau$; thus, $\alpha_k=-(2\alpha_\tau+1)$.

In Figure~\ref{fig:spectra}, $\langle\Delta B_{tot}/B\rangle$ and $\langle\Delta B'\!/B\rangle$ scale identically for each cloud at $\tau\lesssim10^3\,\text{s}$. At $10^1\,\text{s}\lesssim \tau\lesssim10^2\,\text{s}$, $\alpha_\tau=0.35$ in cloud 1 and $\alpha_\tau=0.31$ in cloud 2, equivalent to $\alpha_k=-1.70$ and $\alpha_k=-1.62$, respectively. These values are close to the well known and widely observed $-5/3$ index that features in various theories of MHD turbulence. The scalings of $\langle\Delta B_{tot}/B\rangle$ and $\langle\Delta B'\!/B\rangle$ diverge at $\tau\gtrsim10^3\,\text{s}$ in cloud 1 and at a slightly larger timescale in cloud 2. This divergence reflects the dominance of the flux rope at large scales. At $\tau\gtrsim10^4\,\text{s}$, $\langle\Delta B_{tot}/B\rangle$ scales with $\alpha_\tau=0.80$ in cloud 1 and $\alpha_\tau=0.46$ in cloud 2, equivalent to $\alpha_k=-2.60$ and $\alpha_k=-1.92$, respectively. These relatively steep spectral indices at large scales are due to the flux rope fields; at $\tau\lesssim10^{4}$, the flux rope fits both have $\Delta B_{FR}/B_{FR}$ increments that scale with $\alpha_\tau=1$ (gold lines), equivalent to $\alpha_k=-3$. The scaling is relatively flat at $\tau\gtrsim\tau'_c$ in the rope-subtracted fields.

\begin{figure}
\centerline{\includegraphics[width=0.6\textwidth,clip=]{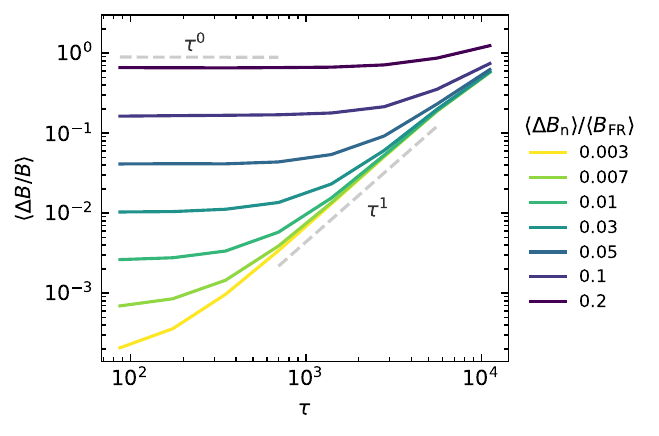}}
\small
        \caption{Scaling of mean normalised increments in the Gold-Hoyle rope fit made to cloud 2 with the addition of $1/f$ noise at different amplitudes.}
\label{fig:noise}
\end{figure}

It can be seen in Figure~\ref{fig:spectra} that the $\langle\Delta B'\!/B\rangle$ fluctuations relative to the flux rope amplitude (i.e. the ratio of the black to gold lines) is significantly smaller in cloud 1 than cloud 2 at all scales. One effect of this reduced ratio is that the flux rope scaling in the total field is less obscured in cloud 1 at large scales. This effect is illustrated in Figure~\ref{fig:noise}, which shows the increment scaling of the Gold-Hoyle fit to cloud 2 with $1/f$ noise added to the profile at a range of different amplitudes. As the noise amplitude relative to the flux rope amplitude increases, there is a transition from the $\tau^1\equiv k^{-3}$ power-law scaling of the flux rope to the $\tau^0\equiv k^{-1}$ scaling of the noise.

\section{Discussion} 
    \label{sec:discussion} 

Through the analysis of two example events, we have illustrated key aspects relating to the measurement of correlation lengths in magnetic clouds. The correlation length may be interpreted as the characteristic scale of the largest `coherent' or correlated structure manifested in a time series of spacecraft measurements. In solar wind intervals without significant background trends, correlation length measurements will directly relate to the outer scale of the turbulent cascade. In magnetic cloud intervals, the largest correlated structure will be the flux rope rather than any turbulent structuring, and correlation measurements will not generally relate to the outer scale unless the flux rope trend is first removed.

The flux rope in a magnetic cloud provides an example of a reasonably well understood trend that can be parametrised with physics-based fitting models. As we have highlighted, even relatively simple flux ropes show a diversity of forms; the large-angle rotation in cloud 1 displayed a Lundquist rope geometry while the shallower rotation of cloud 2 had a Gold-Hoyle structuring. Many other flux rope fitting models of varying complexity have been proposed. There is a degree of subjectivity in choosing which fitting model is appropriate for a particular flux rope interval and correlation length values will have some sensitivity to this choice. Low-order polynomial fits made separately to the three field components could provide a more straightforward, model-independent way to capture the flux rope trend. Polynomial detrending would likely give similar correlation lengths to detrending with a flux rope model but would lack the physical constraints provided by the latter (e.g. the force-free condition). Correlations in magnetic clouds at different scales could also be explored with high and low-pass frequency filtering of data, similar to the analysis of ICME sheaths performed by \citet{Alalahti20}. In practice, sensitivity to the detrending method and also to the duration of the interval analysed \citep{Isaacs15} mean that turbulence correlation lengths cannot be determined in an absolute sense. However, correlation length estimates can be productively analysed in relative terms (e.g. comparing correlation lengths with and without detrending), which has been our primary focus here. Autocorrelation curves calculated with alternative definitions of $C(\tau)$ with different normalisation factors (e.g. $\left\langle(|\delta\bm{B}(t)|+|\delta\bm{B}(t+\tau)|)^2\right\rangle$/4 rather than $\left\langle|\delta\bm{B}(t)|^2\right\rangle)$ were tested, but differences with the curves obtained from Equation~\ref{eqn:autocorr} were negligible.

Increments in the fitted flux rope fields scaled equivalently to a $k^{-3}$ power spectrum. Without detrending, the rope rotation contributes most of the power found at the lowest $k$ values in Fourier spectra of magnetic field fluctuations in cloud intervals \citep{Good23}, and the relative steepness seen in the spectra at low $k$ is due to the $k^{-3}$ scaling of the rope. As $k$ increases, the steep drop-off in power of the flux rope field leaves the turbulent fluctuations as the dominant contribution to the total fluctuation power: at scales in the inertial range, variations in the flux rope field are negligibly small and detrending is not required to analyse the turbulence spectrum. The $k^{-3}$ scaling associated with flux ropes contrasts with the $k^{-2}$ scaling of linear trends, i.e. ramp functions \citep{Matthaeus82}. In an early study of ICME turbulence, \citet{Ruzmaikin97} suggested that steep spectral slopes at large scales in ICME intervals are due to the presence of small-scale discontinuities; however, we have found that these steep slopes are due to the flux rope field.

It has been assumed that the correlation lengths measured from the fit-detrended time series related to the turbulence outer scale. However, if a component of the flux rope field remained after detrending -- e.g., some second-order features of the ropes not captured by the Lundquist and Gold-Hoyle fits, at intermediate scales  between the fit scale and the turbulence spectrum (i.e. at `mesoscales') -- the correlation length measurements could have been sensitive to this component and hence not represent the outer scale; such a component would not act as an energy source for the turbulent cascade given that it would be part of the relatively static magnetic structure of the flux rope. There may also be non-turbulent mesoscale fluctuations unrelated to the rope structure that do inject energy to the cascade. \citet{Good23} found that mesoscale fluctuations with a $k^{-1}$ scaling were present in some of the magnetic clouds they examined; the nature of these mesoscale fluctuations are the subject of ongoing investigations. Furthermore, an analysis of turbulence correlation lengths in magnetic clouds (and ICMEs in general) at 1~au following the approach outlined in this work could be performed; knowledge of these scale lengths, which will be inherently shorter than the previously obtained scale lengths of global ICME structure \citep[e.g.][]{Lugaz18}, would provide new insights for the effective positioning of space weather monitors upstream of the Earth.

\section{Conclusion} 
    \label{sec:conclusion} 

Magnetic field time series measured in magnetic clouds display a flux rope structure at global scales and turbulent fluctuations at smaller scales. With removal of the flux rope trend from the time series, correlation length measurements will give more accurate estimates of the turbulence outer scale. Without detrending, correlation lengths will be sensitive to the scale of the flux rope, with this sensitivity being greater when fluctuation amplitudes are small relative to the flux rope amplitude at the outer scale. The correlation lengths measured from observations of two magnetic clouds detrended with force-free flux rope fits were considerably shorter than the lengths measured without detrending. In addition, flux ropes can significantly steepen the total fluctuation power spectrum at the largest measurable scales, with the flux rope field itself having increments that scale equivalently to a $k^{-3}$ power spectrum.

\begin{acks}
SWG thanks No\'e Lugaz for useful discussions at the American Geophysical Union Fall Meeting, 2024. JLP thanks the Erasmus+ programme for supporting an exchange visit during February--June 2024 to the University of Helsinki.
\end{acks}

\begin{authorcontribution}
SWG wrote the manuscript, performed the analysis described in Section~\ref{sec:rope_fits} and \ref{sec:corr_length}, and produced Figure~\ref{fig:correlations}. JLP performed the analysis described in Section~\ref{sec:B_rot} and produced Figures~\ref{fig:events} and \ref{fig:distributions}--\ref{fig:noise}. A-SMJ and EKJK contributed to the interpretation of results and preparation of the manuscript.
\end{authorcontribution}

\begin{fundinginformation}
This work was funded by an Academy Fellowship awarded by the Research Council of Finland (grants 338486, 346612 and 359914; INERTUM). EKJK acknowledges support from the Research Council of Finland's Centre of Excellence in Research of Sustainable Space (grant 352850; FORESAIL). 
\end{fundinginformation}

\begin{dataavailability}
Spacecraft data used in the analysis were obtained from the CDAWeb archive at https://cdaweb.gsfc.nasa.gov.
\end{dataavailability}

\begin{ethics}
\begin{conflict}
The authors declare that they have no conflicts of interest.
\end{conflict}
\end{ethics}
  

\bibliographystyle{spr-mp-sola-notitles}
\bibliography{references}  

\begin{thebibliography}{43}
\ifx\bisbn     \undefined \def\bisbn  #1{ISBN #1}\fi
\ifx\binits    \undefined \def\binits#1{#1}\fi
\ifx\bauthor   \undefined \def\bauthor#1{#1}\fi
\ifx\batitle   \undefined \def\batitle#1{#1}\fi
\ifx\bjtitle   \undefined \def\bjtitle#1{\textit{#1}}\fi
\ifx\bvolume   \undefined \def\bvolume#1{\textbf{#1}}\fi
\ifx\byear     \undefined \def\byear#1{#1}\fi
\ifx\bissue    \undefined \def\bissue#1{#1}\fi
\ifx\bfpage    \undefined \def\bfpage#1{#1}\fi
\ifx\blpage    \undefined \def\blpage #1{#1}\fi
\ifx\burl      \undefined \def\burl#1{#1}\fi
\ifx\href      \undefined \def\href#1#2{#2}\fi
\ifx\betal     \undefined \def\betal{et al.}\fi
\ifx\bctitle   \undefined \def\bctitle#1{#1}\fi
\ifx\beditor   \undefined \def\beditor#1{#1}\fi
\ifx\bbtitle   \undefined \def\bbtitle#1{\textit{#1}}\fi
\ifx\bedition  \undefined \def\bedition#1{#1}\fi
\ifx\bseriesno \undefined \def\bseriesno#1{\textbf{#1}}\fi
\ifx\blocation \undefined \def\blocation#1{#1}\fi
\ifx\bsertitle \undefined \def\bsertitle#1{\textit{#1}}\fi
\ifx\bsnm      \undefined \def\bsnm#1{#1}\fi
\ifx\bsuffix   \undefined \def\bsuffix#1{#1}\fi
\ifx\bparticle \undefined \def\bparticle#1{#1}\fi
\ifx\barticle  \undefined \def\barticle#1{}\fi
\ifx\binstitute  \undefined \def\binstitute#1{#1}\fi
\ifx\bpublisher  \undefined \def\bpublisher#1{#1}\fi
\ifx\doiurl    \undefined \def\doiurl#1{\href{#1}{DOI}}\fi
\makeatletter
\def\safeHref#1#2#3{\in@{http}{#2}\ifin@\href{#2}{#3}\else\href{#1#2}{#3}\fi}
\makeatother
\ifx\adsurl    \undefined
  \def\adsurl#1{\safeHref{https://ui.adsabs.harvard.edu/abs/}{#1}{ADS}}\fi
\ifx\arxivurl  \undefined
  \def\arxivurl#1{\safeHref{http://arxiv.org/abs/}{#1}{arXiv}}\fi
\ifx\botherref \undefined \def\botherref#1{}\fi
\ifx\url       \undefined \def\url#1{#1}\fi
\ifx\bchapter  \undefined \def\bchapter#1{}\fi
\ifx\bbook     \undefined \def\bbook#1{}\fi
\ifx\bcomment  \undefined \def\bcomment#1{#1}\fi
\ifx\oauthor   \undefined \def\oauthor#1{#1}\fi
\ifx\citeauthoryear \undefined\def \citeauthoryear#1{#1}\fi
\def\endbibitem {}
\ifx\bconflocation  \undefined \def\bconflocation#1{#1} \fi

\bibitem[\protect\citeauthoryear{{Al-Haddad} and {Lugaz}}{2025}]{AlHaddad25}
\begin{barticle}
\bauthor{\bsnm{{Al-Haddad}}, \binits{N.}},
\bauthor{\bsnm{{Lugaz}}, \binits{N.}}:
\byear{2025},
\bjtitle{\ssr}
\bvolume{221},
\bfpage{12}.
\doiurl{https://doi.org/10.1007/s11214-025-01138-w}.
\adsurl{2025SSRv..221...12A}.
\end{barticle}
\endbibitem

\bibitem[\protect\citeauthoryear{{Ala-Lahti} et~al.}{2020}]{Alalahti20}
\begin{barticle}
\bauthor{\bsnm{{Ala-Lahti}}, \binits{M.}},
\bauthor{\bsnm{{Ruohotie}}, \binits{J.}},
\bauthor{\bsnm{{Good}}, \binits{S.}},
\bauthor{\bsnm{{Kilpua}}, \binits{E.K.J.}},
\bauthor{\bsnm{{Lugaz}}, \binits{N.}}:
\byear{2020},
\bjtitle{\jgr}
\bvolume{125},
\bfpage{e2020JA028002}.
\doiurl{https://doi.org/10.1029/2020JA028002}.
\adsurl{2020JGRA..12528002A}.
\end{barticle}
\endbibitem

\bibitem[\protect\citeauthoryear{{Bai} et~al.}{2025}]{Bai25}
\begin{barticle}
\bauthor{\bsnm{{Bai}}, \binits{H.}},
\bauthor{\bsnm{{Huang}}, \binits{S.Y.}},
\bauthor{\bsnm{{Zhang}}, \binits{J.}},
\bauthor{\bsnm{{Yuan}}, \binits{Z.G.}},
\bauthor{\bsnm{{Wu}}, \binits{H.H.}},
\bauthor{\bsnm{{Jiang}}, \binits{K.}},
\bauthor{\bsnm{{Wang}}, \binits{Z.}},
\bauthor{\bsnm{{Xiong}}, \binits{Q.Y.}},
\bauthor{\bsnm{{Lin}}, \binits{R.T.}},
\bauthor{\bsnm{{Tang}}, \binits{Y.T.}}:
\byear{2025},
\bjtitle{\apj}
\bvolume{985},
\bfpage{67}.
\doiurl{https://doi.org/10.3847/1538-4357/adcbae}.
\adsurl{2025ApJ...985...67B}.
\end{barticle}
\endbibitem

\bibitem[\protect\citeauthoryear{{Bale} et~al.}{2016}]{Bale16}
\begin{barticle}
\bauthor{\bsnm{{Bale}}, \binits{S.D.}},
\bauthor{\bsnm{{Goetz}}, \binits{K.}},
\bauthor{\bsnm{{Harvey}}, \binits{P.R.}},
\bauthor{\bsnm{{Turin}}, \binits{P.}},
\bauthor{\bsnm{{Bonnell}}, \binits{J.W.}},
\bauthor{\bsnm{{Dudok de Wit}}, \binits{T.}},
\bauthor{\bsnm{{Ergun}}, \binits{R.E.}},
\bauthor{\bsnm{{MacDowall}}, \binits{R.J.}},
\bauthor{\bsnm{{Pulupa}}, \binits{M.}}, \betal:
\byear{2016},
\bjtitle{\ssr}
\bvolume{204},
\bfpage{49}.
\doiurl{https://doi.org/10.1007/s11214-016-0244-5}.
\adsurl{2016SSRv..204...49B}.
\end{barticle}
\endbibitem

\bibitem[\protect\citeauthoryear{{Banu} et~al.}{2025}]{Banu25}
\begin{barticle}
\bauthor{\bsnm{{Banu}}, \binits{S.A.}},
\bauthor{\bsnm{{Lugaz}}, \binits{N.}},
\bauthor{\bsnm{{Zhuang}}, \binits{B.}},
\bauthor{\bsnm{{Al-Haddad}}, \binits{N.}},
\bauthor{\bsnm{{Farrugia}}, \binits{C.J.}},
\bauthor{\bsnm{{Galvin}}, \binits{A.B.}}:
\byear{2025},
\bjtitle{\apj}
\bvolume{982},
\bfpage{47}.
\doiurl{https://doi.org/10.3847/1538-4357/adb60c}.
\adsurl{2025ApJ...982...47B}.
\end{barticle}
\endbibitem

\bibitem[\protect\citeauthoryear{{Bhattacharjee}
  et~al.}{2023}]{Bhattacharjee23}
\begin{barticle}
\bauthor{\bsnm{{Bhattacharjee}}, \binits{D.}},
\bauthor{\bsnm{{Subramanian}}, \binits{P.}},
\bauthor{\bsnm{{Nieves-Chinchilla}}, \binits{T.}},
\bauthor{\bsnm{{Vourlidas}}, \binits{A.}}:
\byear{2023},
\bjtitle{\mnras}
\bvolume{518},
\bfpage{1185}.
\doiurl{https://doi.org/10.1093/mnras/stac3186}.
\adsurl{2023MNRAS.518.1185B}.
\end{barticle}
\endbibitem

\bibitem[\protect\citeauthoryear{{Borovsky}}{2008}]{Borovsky08}
\begin{barticle}
\bauthor{\bsnm{{Borovsky}}, \binits{J.E.}}:
\byear{2008},
\bjtitle{\jgr}
\bvolume{113},
\bfpage{A08110}.
\doiurl{https://doi.org/10.1029/2007JA012684}.
\adsurl{2008JGRA..113.8110B}.
\end{barticle}
\endbibitem

\bibitem[\protect\citeauthoryear{{Bothmer} and {Schwenn}}{1998}]{Bothmer98}
\begin{barticle}
\bauthor{\bsnm{{Bothmer}}, \binits{V.}},
\bauthor{\bsnm{{Schwenn}}, \binits{R.}}:
\byear{1998},
\bjtitle{\ag}
\bvolume{16},
\bfpage{1}.
\doiurl{https://doi.org/10.1007/s00585-997-0001-x}.
\adsurl{1998AnGeo..16....1B}.
\end{barticle}
\endbibitem

\bibitem[\protect\citeauthoryear{{Burlaga} et~al.}{1981}]{Burlaga81}
\begin{barticle}
\bauthor{\bsnm{{Burlaga}}, \binits{L.}},
\bauthor{\bsnm{{Sittler}}, \binits{E.}},
\bauthor{\bsnm{{Mariani}}, \binits{F.}},
\bauthor{\bsnm{{Schwenn}}, \binits{R.}}:
\byear{1981},
\bjtitle{\jgr}
\bvolume{86},
\bfpage{6673}.
\doiurl{https://doi.org/10.1029/JA086iA08p06673}.
\adsurl{1981JGR....86.6673B}.
\end{barticle}
\endbibitem

\bibitem[\protect\citeauthoryear{{Chen} et~al.}{2015}]{Chen15}
\begin{barticle}
\bauthor{\bsnm{{Chen}}, \binits{C.H.K.}},
\bauthor{\bsnm{{Matteini}}, \binits{L.}},
\bauthor{\bsnm{{Burgess}}, \binits{D.}},
\bauthor{\bsnm{{Horbury}}, \binits{T.S.}}:
\byear{2015},
\bjtitle{\mnras}
\bvolume{453},
\bfpage{L64}.
\doiurl{https://doi.org/10.1093/mnrasl/slv107}.
\adsurl{2015MNRAS.453L..64C}.
\end{barticle}
\endbibitem

\bibitem[\protect\citeauthoryear{{Daei} et~al.}{2023}]{Daei23}
\begin{barticle}
\bauthor{\bsnm{{Daei}}, \binits{F.}},
\bauthor{\bsnm{{Pomoell}}, \binits{J.}},
\bauthor{\bsnm{{Price}}, \binits{D.J.}},
\bauthor{\bsnm{{Kumari}}, \binits{A.}},
\bauthor{\bsnm{{Good}}, \binits{S.}},
\bauthor{\bsnm{{Kilpua}}, \binits{E.K.J.}}:
\byear{2023},
\bjtitle{\aap}
\bvolume{676},
\bfpage{A141}.
\doiurl{https://doi.org/10.1051/0004-6361/202346183}.
\adsurl{2023A&A...676A.141D}.
\end{barticle}
\endbibitem

\bibitem[\protect\citeauthoryear{{Fox} et~al.}{2016}]{Fox16}
\begin{barticle}
\bauthor{\bsnm{{Fox}}, \binits{N.J.}},
\bauthor{\bsnm{{Velli}}, \binits{M.C.}},
\bauthor{\bsnm{{Bale}}, \binits{S.D.}},
\bauthor{\bsnm{{Decker}}, \binits{R.}},
\bauthor{\bsnm{{Driesman}}, \binits{A.}},
\bauthor{\bsnm{{Howard}}, \binits{R.A.}}, \betal:
\byear{2016},
\bjtitle{\ssr}
\bvolume{204},
\bfpage{7}.
\doiurl{https://doi.org/10.1007/s11214-015-0211-6}.
\adsurl{2016SSRv..204....7F}.
\end{barticle}
\endbibitem

\bibitem[\protect\citeauthoryear{{Gold} and {Hoyle}}{1960}]{GoldHoyle60}
\begin{barticle}
\bauthor{\bsnm{{Gold}}, \binits{T.}},
\bauthor{\bsnm{{Hoyle}}, \binits{F.}}:
\byear{1960},
\bjtitle{\mnras}
\bvolume{120},
\bfpage{89}.
\doiurl{https://doi.org/10.1093/mnras/120.2.89}.
\adsurl{1960MNRAS.120...89G}.
\end{barticle}
\endbibitem

\bibitem[\protect\citeauthoryear{{Good} et~al.}{2018}]{Good18}
\begin{barticle}
\bauthor{\bsnm{{Good}}, \binits{S.W.}},
\bauthor{\bsnm{{Forsyth}}, \binits{R.J.}},
\bauthor{\bsnm{{Eastwood}}, \binits{J.P.}},
\bauthor{\bsnm{{M{\"o}stl}}, \binits{C.}}:
\byear{2018},
\bjtitle{\solphys}
\bvolume{293},
\bfpage{52}.
\doiurl{https://doi.org/10.1007/s11207-018-1264-y}.
\adsurl{2018SoPh..293...52G}.
\end{barticle}
\endbibitem

\bibitem[\protect\citeauthoryear{{Good} et~al.}{2019}]{Good19}
\begin{barticle}
\bauthor{\bsnm{{Good}}, \binits{S.W.}},
\bauthor{\bsnm{{Kilpua}}, \binits{E.K.J.}},
\bauthor{\bsnm{{LaMoury}}, \binits{A.T.}},
\bauthor{\bsnm{{Forsyth}}, \binits{R.J.}},
\bauthor{\bsnm{{Eastwood}}, \binits{J.P.}},
\bauthor{\bsnm{{M{\"o}stl}}, \binits{C.}}:
\byear{2019},
\bjtitle{\jgr}
\bvolume{124},
\bfpage{4960}.
\doiurl{https://doi.org/10.1029/2019JA026475}.
\adsurl{2019JGRA..124.4960G}.
\end{barticle}
\endbibitem

\bibitem[\protect\citeauthoryear{{Good} et~al.}{2020a}]{Good20a}
\begin{barticle}
\bauthor{\bsnm{{Good}}, \binits{S.W.}},
\bauthor{\bsnm{{Kilpua}}, \binits{E.K.J.}},
\bauthor{\bsnm{{Ala-Lahti}}, \binits{M.}},
\bauthor{\bsnm{{Osmane}}, \binits{A.}},
\bauthor{\bsnm{{Bale}}, \binits{S.D.}},
\bauthor{\bsnm{{Zhao}}, \binits{L.-L.}}:
\byear{2020}a,
\bjtitle{\apjl}
\bvolume{900},
\bfpage{L32}.
\doiurl{https://doi.org/10.3847/2041-8213/abb021}.
\adsurl{2020ApJ...900L..32G}.
\end{barticle}
\endbibitem

\bibitem[\protect\citeauthoryear{{Good} et~al.}{2020b}]{Good20b}
\begin{barticle}
\bauthor{\bsnm{{Good}}, \binits{S.W.}},
\bauthor{\bsnm{{Ala-Lahti}}, \binits{M.}},
\bauthor{\bsnm{{Palmerio}}, \binits{E.}},
\bauthor{\bsnm{{Kilpua}}, \binits{E.K.J.}},
\bauthor{\bsnm{{Osmane}}, \binits{A.}}:
\byear{2020}b,
\bjtitle{\apj}
\bvolume{893},
\bfpage{110}.
\doiurl{https://doi.org/10.3847/1538-4357/ab7fa2}.
\adsurl{2020ApJ...893..110G}.
\end{barticle}
\endbibitem

\bibitem[\protect\citeauthoryear{{Good} et~al.}{2023}]{Good23}
\begin{barticle}
\bauthor{\bsnm{{Good}}, \binits{S.W.}},
\bauthor{\bsnm{{Rantala}}, \binits{O.K.}},
\bauthor{\bsnm{{Jylh{\"a}}}, \binits{A.-S.M.}},
\bauthor{\bsnm{{Chen}}, \binits{C.H.K.}},
\bauthor{\bsnm{{M{\"o}stl}}, \binits{C.}},
\bauthor{\bsnm{{Kilpua}}, \binits{E.K.J.}}:
\byear{2023},
\bjtitle{\apjl}
\bvolume{956},
\bfpage{L30}.
\doiurl{https://doi.org/10.3847/2041-8213/acfd1c}.
\adsurl{2023ApJ...956L..30G}.
\end{barticle}
\endbibitem

\bibitem[\protect\citeauthoryear{{Isaacs}, {Tessein}, and
  {Matthaeus}}{2015}]{Isaacs15}
\begin{barticle}
\bauthor{\bsnm{{Isaacs}}, \binits{J.J.}},
\bauthor{\bsnm{{Tessein}}, \binits{J.A.}},
\bauthor{\bsnm{{Matthaeus}}, \binits{W.H.}}:
\byear{2015},
\bjtitle{\jgr}
\bvolume{120},
\bfpage{868}.
\doiurl{https://doi.org/10.1002/2014JA020661}.
\adsurl{2015JGRA..120..868I}.
\end{barticle}
\endbibitem

\bibitem[\protect\citeauthoryear{{Khuntia} and {Mishra}}{2025}]{Khuntia25}
\begin{barticle}
\bauthor{\bsnm{{Khuntia}}, \binits{S.}},
\bauthor{\bsnm{{Mishra}}, \binits{W.}}:
\byear{2025},
\bjtitle{J. Astrophys. Astron.}
\bvolume{46},
\bfpage{70}.
\doiurl{https://doi.org/10.1007/s12036-025-10085-5}.
\adsurl{2025JApA...46...70K}.
\end{barticle}
\endbibitem

\bibitem[\protect\citeauthoryear{{Kilpua}, {Koskinen}, and
  {Pulkkinen}}{2017}]{Kilpua17}
\begin{barticle}
\bauthor{\bsnm{{Kilpua}}, \binits{E.}},
\bauthor{\bsnm{{Koskinen}}, \binits{H.E.J.}},
\bauthor{\bsnm{{Pulkkinen}}, \binits{T.I.}}:
\byear{2017},
\bjtitle{\lrsp}
\bvolume{14},
\bfpage{5}.
\doiurl{https://doi.org/10.1007/s41116-017-0009-6}.
\adsurl{2017LRSP...14....5K}.
\end{barticle}
\endbibitem

\bibitem[\protect\citeauthoryear{{Kilpua} et~al.}{2011}]{Kilpua11}
\begin{barticle}
\bauthor{\bsnm{{Kilpua}}, \binits{E.K.J.}},
\bauthor{\bsnm{{Jian}}, \binits{L.K.}},
\bauthor{\bsnm{{Li}}, \binits{Y.}},
\bauthor{\bsnm{{Luhmann}}, \binits{J.G.}},
\bauthor{\bsnm{{Russell}}, \binits{C.T.}}:
\byear{2011},
\bjtitle{\jastp}
\bvolume{73},
\bfpage{1228}.
\doiurl{https://doi.org/10.1016/j.jastp.2010.10.012}.
\adsurl{2011JASTP..73.1228K}.
\end{barticle}
\endbibitem

\bibitem[\protect\citeauthoryear{{Kilpua} et~al.}{2020}]{Kilpua20}
\begin{barticle}
\bauthor{\bsnm{{Kilpua}}, \binits{E.K.J.}},
\bauthor{\bsnm{{Fontaine}}, \binits{D.}},
\bauthor{\bsnm{{Good}}, \binits{S.W.}},
\bauthor{\bsnm{{Ala-Lahti}}, \binits{M.}},
\bauthor{\bsnm{{Osmane}}, \binits{A.}},
\bauthor{\bsnm{{Palmerio}}, \binits{E.}},
\bauthor{\bsnm{{Yordanova}}, \binits{E.}},
\bauthor{\bsnm{{Moissard}}, \binits{C.}},
\bauthor{\bsnm{{Hadid}}, \binits{L.Z.}},
\bauthor{\bsnm{{Janvier}}, \binits{M.}}:
\byear{2020},
\bjtitle{\ag}
\bvolume{38},
\bfpage{999}.
\doiurl{https://doi.org/10.5194/angeo-38-999-2020}.
\adsurl{2020AnGeo..38..999K}.
\end{barticle}
\endbibitem

\bibitem[\protect\citeauthoryear{{Larosa} et~al.}{2024}]{Larosa24}
\begin{barticle}
\bauthor{\bsnm{{Larosa}}, \binits{A.}},
\bauthor{\bsnm{{Chen}}, \binits{C.H.K.}},
\bauthor{\bsnm{{McIntyre}}, \binits{J.R.}},
\bauthor{\bsnm{{Jagarlamudi}}, \binits{V.K.}},
\bauthor{\bsnm{{Sorriso-Valvo}}, \binits{L.}}:
\byear{2024},
\bjtitle{\aap}
\bvolume{686},
\bfpage{A238}.
\doiurl{https://doi.org/10.1051/0004-6361/202450030}.
\adsurl{2024A&A...686A.238L}.
\end{barticle}
\endbibitem

\bibitem[\protect\citeauthoryear{{Lugaz} et~al.}{2018}]{Lugaz18}
\begin{barticle}
\bauthor{\bsnm{{Lugaz}}, \binits{N.}},
\bauthor{\bsnm{{Farrugia}}, \binits{C.J.}},
\bauthor{\bsnm{{Winslow}}, \binits{R.M.}},
\bauthor{\bsnm{{Al-Haddad}}, \binits{N.}},
\bauthor{\bsnm{{Galvin}}, \binits{A.B.}},
\bauthor{\bsnm{{Nieves-Chinchilla}}, \binits{T.}}, \betal:
\byear{2018},
\bjtitle{\apjl}
\bvolume{864},
\bfpage{L7}.
\doiurl{https://doi.org/10.3847/2041-8213/aad9f4}.
\adsurl{2018ApJ...864L...7L}.
\end{barticle}
\endbibitem

\bibitem[\protect\citeauthoryear{{Lugaz} et~al.}{2025}]{Lugaz25}
\begin{barticle}
\bauthor{\bsnm{{Lugaz}}, \binits{N.}},
\bauthor{\bsnm{{Al-Haddad}}, \binits{N.}},
\bauthor{\bsnm{{Zhuang}}, \binits{B.}},
\bauthor{\bsnm{{M{\"o}stl}}, \binits{C.}},
\bauthor{\bsnm{{Davies}}, \binits{E.E.}},
\bauthor{\bsnm{{Farrugia}}, \binits{C.J.}}, \betal:
\byear{2025},
\bjtitle{Space Weather}
\bvolume{23},
\bfpage{2024SW004189}.
\doiurl{https://doi.org/10.1029/2024SW004189}.
\adsurl{2025SpWea..2304189L}.
\end{barticle}
\endbibitem

\bibitem[\protect\citeauthoryear{{Lundquist}}{1951}]{Lundquist51}
\begin{barticle}
\bauthor{\bsnm{{Lundquist}}, \binits{S.}}:
\byear{1951},
\bjtitle{Phys. Rev.}
\bvolume{83},
\bfpage{307}.
\doiurl{https://doi.org/10.1103/PhysRev.83.307}.
\adsurl{1951PhRv...83..307L}.
\end{barticle}
\endbibitem

\bibitem[\protect\citeauthoryear{{Matteini} et~al.}{2018}]{Matteini18}
\begin{barticle}
\bauthor{\bsnm{{Matteini}}, \binits{L.}},
\bauthor{\bsnm{{Stansby}}, \binits{D.}},
\bauthor{\bsnm{{Horbury}}, \binits{T.S.}},
\bauthor{\bsnm{{Chen}}, \binits{C.H.K.}}:
\byear{2018},
\bjtitle{\apjl}
\bvolume{869},
\bfpage{L32}.
\doiurl{https://doi.org/10.3847/2041-8213/aaf573}.
\adsurl{2018ApJ...869L..32M}.
\end{barticle}
\endbibitem

\bibitem[\protect\citeauthoryear{{Matthaeus} and
  {Goldstein}}{1982}]{Matthaeus82}
\begin{barticle}
\bauthor{\bsnm{{Matthaeus}}, \binits{W.H.}},
\bauthor{\bsnm{{Goldstein}}, \binits{M.L.}}:
\byear{1982},
\bjtitle{\jgr}
\bvolume{87},
\bfpage{6011}.
\doiurl{https://doi.org/10.1029/JA087iA08p06011}.
\adsurl{1982JGR....87.6011M}.
\end{barticle}
\endbibitem

\bibitem[\protect\citeauthoryear{{Monin} and {Yaglom}}{1975}]{Monin75}
\begin{bbook}
\bauthor{\bsnm{{Monin}}, \binits{A.S.}},
\bauthor{\bsnm{{Yaglom}}, \binits{A.M.}}:
\byear{1975},
\bbtitle{{Statistical Fluid Mechanics: Mechanics of Turbulence. Volume 2}},
\bpublisher{{MIT Press}},
\blocation{{Cambridge, Mass.}}
\adsurl{1975mit..bookR....M}.
\end{bbook}
\endbibitem

\bibitem[\protect\citeauthoryear{{Owens}}{2020}]{Owens20}
\begin{barticle}
\bauthor{\bsnm{{Owens}}, \binits{M.J.}}:
\byear{2020},
\bjtitle{\solphys}
\bvolume{295},
\bfpage{148}.
\doiurl{https://doi.org/10.1007/s11207-020-01721-0}.
\adsurl{2020SoPh..295..148O}.
\end{barticle}
\endbibitem

\bibitem[\protect\citeauthoryear{{Palmerio}}{2025}]{Palmerio25b}
\begin{barticle}
\bauthor{\bsnm{{Palmerio}}, \binits{E.}}:
\byear{2025},
\bjtitle{Space Weather}
\bvolume{23},
\bfpage{e2025SW004452}.
\doiurl{https://doi.org/10.1029/2025SW004452}.
\adsurl{2025SpWea..2304452P}.
\end{barticle}
\endbibitem

\bibitem[\protect\citeauthoryear{{Rakhmanova} et~al.}{2025}]{Rakhmanova25}
\begin{barticle}
\bauthor{\bsnm{{Rakhmanova}}, \binits{L.}},
\bauthor{\bsnm{{Riazantseva}}, \binits{M.}},
\bauthor{\bsnm{{Yermolaev}}, \binits{Y.}},
\bauthor{\bsnm{{Khokhlachev}}, \binits{A.}},
\bauthor{\bsnm{{Zastenker}}, \binits{G.}}:
\byear{2025},
\bjtitle{\solphys}
\bvolume{300},
\bfpage{44}.
\doiurl{https://doi.org/10.1007/s11207-025-02458-4}.
\adsurl{2025SoPh..300...44R}.
\end{barticle}
\endbibitem

\bibitem[\protect\citeauthoryear{{Ruiz} et~al.}{2014}]{Ruiz14}
\begin{barticle}
\bauthor{\bsnm{{Ruiz}}, \binits{M.E.}},
\bauthor{\bsnm{{Dasso}}, \binits{S.}},
\bauthor{\bsnm{{Matthaeus}}, \binits{W.H.}},
\bauthor{\bsnm{{Weygand}}, \binits{J.M.}}:
\byear{2014},
\bjtitle{\solphys}
\bvolume{289},
\bfpage{3917}.
\doiurl{https://doi.org/10.1007/s11207-014-0531-9}.
\adsurl{2014SoPh..289.3917R}.
\end{barticle}
\endbibitem

\bibitem[\protect\citeauthoryear{{Ruohotie} et~al.}{2025}]{Ruohotie25}
\begin{barticle}
\bauthor{\bsnm{{Ruohotie}}, \binits{J.}},
\bauthor{\bsnm{{Good}}, \binits{S.}},
\bauthor{\bsnm{{M{\"o}stl}}, \binits{C.}},
\bauthor{\bsnm{{Kilpua}}, \binits{E.}}:
\byear{2025},
\bjtitle{\apjl}
\bvolume{986},
\bfpage{L27}.
\doiurl{https://doi.org/10.3847/2041-8213/ade0b0}.
\adsurl{2025ApJ...986L..27R}.
\end{barticle}
\endbibitem

\bibitem[\protect\citeauthoryear{{Ruzmaikin}, {Feynman}, and
  {Smith}}{1997}]{Ruzmaikin97}
\begin{barticle}
\bauthor{\bsnm{{Ruzmaikin}}, \binits{A.}},
\bauthor{\bsnm{{Feynman}}, \binits{J.}},
\bauthor{\bsnm{{Smith}}, \binits{E.J.}}:
\byear{1997},
\bjtitle{\jgr}
\bvolume{102},
\bfpage{19753}.
\doiurl{https://doi.org/10.1029/97JA01558}.
\adsurl{1997JGR...10219753R}.
\end{barticle}
\endbibitem

\bibitem[\protect\citeauthoryear{{Scolini} et~al.}{2022}]{Scolini22}
\begin{barticle}
\bauthor{\bsnm{{Scolini}}, \binits{C.}},
\bauthor{\bsnm{{Winslow}}, \binits{R.M.}},
\bauthor{\bsnm{{Lugaz}}, \binits{N.}},
\bauthor{\bsnm{{Salman}}, \binits{T.M.}},
\bauthor{\bsnm{{Davies}}, \binits{E.E.}},
\bauthor{\bsnm{{Galvin}}, \binits{A.B.}}:
\byear{2022},
\bjtitle{\apj}
\bvolume{927},
\bfpage{102}.
\doiurl{https://doi.org/10.3847/1538-4357/ac3e60}.
\adsurl{2022ApJ...927..102S}.
\end{barticle}
\endbibitem

\bibitem[\protect\citeauthoryear{{Shaikh}}{2024}]{Shaikh24}
\begin{barticle}
\bauthor{\bsnm{{Shaikh}}, \binits{Z.I.}}:
\byear{2024},
\bjtitle{\mnras}
\bvolume{530},
\bfpage{3005}.
\doiurl{https://doi.org/10.1093/mnras/stae897}.
\adsurl{2024MNRAS.530.3005S}.
\end{barticle}
\endbibitem

\bibitem[\protect\citeauthoryear{{Weiler} et~al.}{2025}]{Weiler25}
\begin{barticle}
\bauthor{\bsnm{{Weiler}}, \binits{E.}},
\bauthor{\bsnm{{M{\"o}stl}}, \binits{C.}},
\bauthor{\bsnm{{Davies}}, \binits{E.E.}},
\bauthor{\bsnm{{Veronig}}, \binits{A.M.}},
\bauthor{\bsnm{{Amerstorfer}}, \binits{U.V.}},
\bauthor{\bsnm{{Amerstorfer}}, \binits{T.}},
\bauthor{\bsnm{{Le Lou{\"e}dec}}, \binits{J.}},
\bauthor{\bsnm{{Bauer}}, \binits{M.}},
\bauthor{\bsnm{{Lugaz}}, \binits{N.}},
\bauthor{\bsnm{{Haberle}}, \binits{V.}},
\bauthor{\bsnm{{R{\"u}disser}}, \binits{H.T.}},
\bauthor{\bsnm{{Majumdar}}, \binits{S.}},
\bauthor{\bsnm{{Reiss}}, \binits{M.}}:
\byear{2025},
\bjtitle{Space Weather}
\bvolume{23},
\bfpage{2024SW004260}.
\doiurl{https://doi.org/10.1029/2024SW004260}.
\adsurl{2025SpWea..2304260W}.
\end{barticle}
\endbibitem

\bibitem[\protect\citeauthoryear{{Wicks}, {Owens}, and
  {Horbury}}{2010}]{Wicks10}
\begin{barticle}
\bauthor{\bsnm{{Wicks}}, \binits{R.T.}},
\bauthor{\bsnm{{Owens}}, \binits{M.J.}},
\bauthor{\bsnm{{Horbury}}, \binits{T.S.}}:
\byear{2010},
\bjtitle{\solphys}
\bvolume{262},
\bfpage{191}.
\doiurl{https://doi.org/10.1007/s11207-010-9509-4}.
\adsurl{2010SoPh..262..191W}.
\end{barticle}
\endbibitem

\bibitem[\protect\citeauthoryear{{Zhang} et~al.}{2007}]{Zhang07}
\begin{barticle}
\bauthor{\bsnm{{Zhang}}, \binits{J.}},
\bauthor{\bsnm{{Richardson}}, \binits{I.G.}},
\bauthor{\bsnm{{Webb}}, \binits{D.F.}},
\bauthor{\bsnm{{Gopalswamy}}, \binits{N.}},
\bauthor{\bsnm{{Huttunen}}, \binits{E.}},
\bauthor{\bsnm{{Kasper}}, \binits{J.C.}},
\bauthor{\bsnm{{Nitta}}, \binits{N.V.}},
\bauthor{\bsnm{{Poomvises}}, \binits{W.}},
\bauthor{\bsnm{{Thompson}}, \binits{B.J.}},
\bauthor{\bsnm{{Wu}}, \binits{C.-C.}},
\bauthor{\bsnm{{Yashiro}}, \binits{S.}},
\bauthor{\bsnm{{Zhukov}}, \binits{A.N.}}:
\byear{2007},
\bjtitle{\jgr}
\bvolume{112},
\bfpage{A10102}.
\doiurl{https://doi.org/10.1029/2007JA012321}.
\adsurl{2007JGRA..11210102Z}.
\end{barticle}
\endbibitem

\bibitem[\protect\citeauthoryear{{Zhdankin}, {Boldyrev}, and
  {Mason}}{2012}]{Zhdankin12}
\begin{barticle}
\bauthor{\bsnm{{Zhdankin}}, \binits{V.}},
\bauthor{\bsnm{{Boldyrev}}, \binits{S.}},
\bauthor{\bsnm{{Mason}}, \binits{J.}}:
\byear{2012},
\bjtitle{\apjl}
\bvolume{760},
\bfpage{L22}.
\doiurl{https://doi.org/10.1088/2041-8205/760/2/L22}.
\adsurl{2012ApJ...760L..22Z}.
\end{barticle}
\endbibitem

\bibitem[\protect\citeauthoryear{{Zurbuchen} and
  {Richardson}}{2006}]{Zurbuchen06}
\begin{barticle}
\bauthor{\bsnm{{Zurbuchen}}, \binits{T.H.}},
\bauthor{\bsnm{{Richardson}}, \binits{I.G.}}:
\byear{2006},
\bjtitle{\ssr}
\bvolume{123},
\bfpage{31}.
\doiurl{https://doi.org/10.1007/s11214-006-9010-4}.
\adsurl{2006SSRv..123...31Z}.
\end{barticle}
\endbibitem

\end{thebibliography}

\end{document}